# How Can We Avert Dangerous Climate Change?

## James Hansen

*Abstract.* Recent analyses indicate that the amount of atmospheric $CO_2$ required to cause dangerous climate change is at most 450 ppm, and likely less than that. Reductions of non-$CO_2$ climate forcings can provide only moderate, albeit important, adjustments to the $CO_2$ limit. Realization of how close the planet is to 'tipping points' with unacceptable consequences, especially ice sheet disintegration with sea level rise out of humanity's control, has a bright side. It implies an imperative: we must find a way to keep the $CO_2$ amount so low that it will also avert other detrimental effects that had begun to seem inevitable, e.g., ocean acidification, loss of most alpine glaciers and thus the water supply for millions of people, and shifting of climatic zones with consequent extermination of species.

Here I outline from a scientific perspective actions needed to achieve low limits on $CO_2$ and global warming. These changes are technically feasible and have ancillary benefits. Achievement of needed changes requires overcoming the spurious argument that developed and developing countries have equivalent responsibilities, as well as overcoming special interests advocating minimalist or counterproductive actions such as corn-based ethanol and liquid-fuel-from-coal programs.

This paper consists of written testimony that I delivered as a private citizen to the Select Committee on Energy Independence and Global Warming, United States House of Representatives on 26 April 2007. I have added to that testimony: this abstract, references for several statements in the testimony, and some specificity in the final section on solutions.

## 1. Summary

Crystallizing scientific data and analyses reveal that the Earth is close to dangerous climate change, to tipping points of the system with potential for irreversible deleterious effects. This information derives in part from paleoclimate data, i.e., the record of how climate changed in the past, as well as from measurements being made now by satellites and in the field.

The Earth's history shows that climate is remarkably sensitive to global forcings. Positive feedbacks predominate. This has allowed the entire planet to be whipsawed between climate states. Huge natural climate changes, from glacial to interglacial states, have been driven by very weak, very slow forcings, and positive feedbacks.

Now humans are applying a far stronger forcing much more rapidly, as we put back into the atmosphere, in a geologic heartbeat, fossil fuels that accumulated over millions of years. Positive feedbacks are beginning to occur, on a range of time scales.

The climate system has inertia. Nearly full response to a climate forcing requires decades to centuries. But that inertia is not our friend. It means that there is additional climate change in the pipeline that will occur in coming decades even without additional greenhouse gases.

The upshot is that very little additional forcing is needed to cause dramatic effects. To cause the loss of all summer Arctic ice with devastating effects on wildlife and indigenous people. To cause an intensification of subtropical conditions that would greatly exacerbate water shortages in the American West and many other parts of the world, and likely render the semi-arid states from west and central Texas through Oklahoma, Kansas, Nebraska and the Dakotas increasingly drought prone and unsuitable for agriculture. To cause the extermination of a large fraction of plant and animal species, an indictment of humanity's failure to preserve creation.

For humanity itself, the greatest threat is the likely demise of the West Antarctic ice sheet as it is attacked from below by a warming ocean and above by increased surface melt. There is increasing realization that sea level rise this century may be measured in meters if we follow business-as-usual fossil fuel emissions.

There is a bright side to this planetary emergency. We can successfully address the emergency only by stabilizing climate close to its present state; there is no viable option, as adaptation to a continually rising sea level is not practical. Therefore, if we address the problem, there will be no need to adapt to the highly deleterious regional climate changes mentioned above, acidification of the ocean, and other detrimental effects. The actions needed to stabilize climate will preserve creation and restore a cleaner, healthier atmosphere.

The dangerous level of $CO_2$ is at most 450 ppm, and it is probably less. The low limit on $CO_2$ forces us to move promptly to the next phase of the industrial revolution. Changing light bulbs and making ethanol from corn will not solve the problem, although the former act is useful. Science provides a clear outline for what must be done, a four point strategy:

First, we must phase out the use of coal and unconventional fossil fuels except where the $CO_2$ is captured and sequestered. There should be a moratorium on construction of old-technology coal-fired power plants.

Second, there must be a rising price (tax) on carbon emissions, as well as effective energy efficiency standards, and removal of barriers to efficiency. These actions are needed to spur innovation in energy efficiency and renewable energies, and thus to stretch oil and gas supplies to cover the need for mobile fuels during the transition to the next phase of the industrial revolution 'beyond petroleum'.

Third, there should be focused efforts to reduce non-$CO_2$ human-made climate forcings, especially methane, ozone and black carbon.

Fourth, steps must be taken to 'draw down' atmospheric $CO_2$ via improved farming and forestry practices, including burning of biofuels in power plants with $CO_2$ sequestration.

Note that I do not specify an exact fraction by which $CO_2$ emissions must be reduced by 2050 or any other date. Indeed, science is not able to specify an exact requirement now, but we can say that emissions must be reduced to a fraction of their current values. Given the fact that readily available oil will surely be employed for mobile sources, and given the magnitudes of the different fossil fuel reservoirs, it seems best to frame the problem as I have in this four-point strategy, and adjust specific targets and policies as knowledge improves.

Responsibility of the United States for global climate change exceeds that of any other nation by more than a factor of three, even though China is now passing the United States in current emissions. The United States will continue to be primarily responsible for climate change for decades to come.

The above conclusions follow from the science. I go further here in expressing my opinion about the implications of this research for citizens in our democratic system. I believe that this is appropriate, in part because of resistance that the scientific conclusions have met among special interests, and because of misinformation about the science that has been spread. My opinions carry no more weight than those of any other citizen, but conceivably my experience in presenting this research in different circles allows some insight. In any case, I have as much right to express my opinion as do the special interests.

In my opinion, the United States should recognize openly its leading role in causing human-made climate change and promptly take a leadership role in addressing the matter. We have a moral responsibility to do so.

Moreover, it is in our interest to take actions now. We can benefit economically from extensive technology development, with many good high-tech high-pay jobs. Of course, moving to the next phase of the industrial revolution will require changes, dislocations, sacrifices and hard work.  But these provide no basis for inaction.

We cannot let the pleadings and misinformation of special interests determine our actions, special interests driven by motives of short-term profit.  And we cannot shrink from our personal responsibilities.  We are now, through our government, standing alongside the polluters, officially as a hulking 'friend of the court', arguing against limitations on emissions.

Is this the picture of our generation we will leave for our children, a picture of ignorance and greed?  We live in a democracy.  Policies represent our collective will.  We cannot blame others.  If we allow the planet to pass tipping points, to set in motion irreversible changes to the detriment of nature and humanity, it will be hard to explain our role to our children and grandchildren.

We cannot claim, with legitimacy, that 'we did not know'.  In my opinion, it is time for the public to demand, from government and industry, priority for actions needed to preserve the planet for future generations.

## 2. Basis for Testimony

My testimony is derived primarily from the six publications listed below.  It is based on a much broader body of knowledge of the scientific community, which is not practical to document in the brief hours available to prepare this testimony.

The first four publications below are published or 'in press' in regular peer-reviewed scientific journals, each having been reviewed by either two or three scientific peers.  The fifth article is my attempt to describe conclusions from this research in a language intended for a broader audience.  The sixth article has been submitted to a peer-reviewed scientific journal.

A. Dangerous human-made interference with climate: a GISS modelE study
>   Hansen and 46 co-authors (2007), *Atmos. Chem. Phys.*, 7, 2287-2312. PDF available at http://pubs.giss.nasa.gov/abstracts/2007/Hansen_etal_1.html

B. Climate change and trace gases
>   Hansen and 5 co-authors (2007), *Phil. Trans. Royal Soc. A*, 365, 1925-1954, doi:10.1098/rsta.2007.2052. PDF available at http://pubs.giss.nasa.gov/abstracts/2007/Hansen_etal_2.html

C. Climate simulations for 1880-2003 with GISS modelE
>   Hansen and 46 co-authors, in press at *Climate Dynamics*, doi:10.1007/s00382-007-0255-8. PDF available at http://pubs.giss.nasa.gov/abstracts/inpress/Hansen_etal_1.html

D. Scientific reticence and sea level rise
>   Hansen (2007), *Environ. Res. Lett.,* 2, 024002, doi:10.1088/1748-9326/2/2/024002. PDF available at http://pubs.giss.nasa.gov/abstracts/2007/Hansen.html

E. State of the Wild: Perspective of a Climatologist
>   (accepted, to be edited) http://www.giss.nasa.gov/~jhansen/preprints/Wild.070410.pdf

F. Implications of "peak oil" for atmospheric $CO_2$ and climate
>   Kharecha and Hansen, submitted to *Environ. Res. Lett*. PDF available at http://pubs.giss.nasa.gov/abstracts/submitted/Kharecha_Hansen.html

## 3. Crystallizing Science

In the past few years it has become clear that the Earth is close to dangerous climate change, to tipping points of the system with the potential for irreversible deleterious effects.

Paleoclimate data show that climate is remarkably sensitive to global forcings. Positive feedbacks have caused the entire planet to be whipsawed between climate states, driven by very weak climate forcings.

The time scale for full glacial-to-interglacial climate changes is millennia. However, this millennial time scale reflects the time scale of the slow weak climate forcing due to Earth orbital changes, not an inherent climate response time. Indeed, the response time of the climate system to rapid forcings, such as human-made greenhouse gases, will be decades to centuries, a function of ocean mixing time and climate feedbacks.

This decade-century climate response time is unfortunate for humanity. It is long enough to prevent people from seeing immediate consequences of human-made climate forcings, as much of the climate change is still 'in the pipeline'. Yet it is short enough for large climate impacts to occur this century.

The concept of additional global warming 'in the pipeline' is not new, but it has become more ominous through the realization that several nominally 'slow' climate feedbacks are likely to have significant effects on decadal time scales. These include poleward movement of vegetation, darkening and disintegration of ice sheets, and greenhouse gas feedbacks. These 'slow' feedbacks, which are not included in their entirety in standard IPCC simulations, are positive and thus they amplify expected anthropogenic climate change.

The implication of the crystallizing scientific understanding is that the planet is on the verge of dramatic climate change. It is still possible to avoid the most deleterious effects, but only if prompt actions are taken to stabilize global temperature close to its present value. Because of the profound implications, it is appropriate to clarify the basis of these conclusions.

We first discuss fundamental aspects of the climate system: climate forcings, feedbacks and response times. We then make note of how the Earth's climate responded to forcings in the past few million years. Finally, we summarize the basis for the conclusion that present climate is on the verge of critical tipping points.

**A. Climate System**

Climate is an average of weather over some period, including the variability and extremes within that period. Because day-to-day weather fluctuations are so large, it is not easy to notice small changes of the average weather or climate. However, moderate changes of climate can have significant effects, for example, on the ability of plants and animals to survive in a given region and on the stability of large ice masses and thus sea level.

Climate varies a lot without any help from humans. In part the variations are simply chaotic fluctuations of a complex dynamical system, as the atmosphere and ocean are always sloshing about. The climate also responds to natural forcings, such as changes of the brightness of the sun or eruptions of large volcanoes, which discharge small particles into the upper atmosphere where they reflect sunlight and cool the Earth.

*Climate forcing.* A climate forcing is a perturbation of the Earth's energy balance that tends to alter the Earth's temperature (Hansen et al. 2005a). For example, if the sun's brightness increases 2% that is a positive forcing of about 4.5 W/m$^2$ (watts per square meter), as it results in an increase of that amount in the energy absorbed by the Earth. Such a forcing would upset the normal balance that exists between the amount of solar energy absorbed by the Earth and the amount of heat radiation emitted to space by the Earth. The Earth responds to this forcing by warming up until its thermal radiation to space equals the energy absorbed from the sun.

Doubling the amount of carbon dioxide ($CO_2$) in the atmosphere causes a global climate forcing similar in magnitude to that for a 2% increase of solar irradiance. The $CO_2$ forcing works by making the atmosphere more opaque to infrared radiation, the wavelengths of the Earth's heat radiation. As a result of this increased opacity the heat radiation to space arises from greater heights in the atmosphere. Because the temperature falls off with height in the lower atmosphere, energy radiated to space with doubled $CO_2$ is reduced by an amount that is readily calculated from radiation physics to be approximately 4 $W/m^2$. So the planet's energy imbalance is about the same as for a 2% increase of solar irradiance. In either case, the Earth responds by warming up enough to restore energy balance.

Climate models show that, as might be expected, two forcings of similar magnitude yield similar global temperature change, although variations in the "efficacy" of specific forcings of the order of 20% are not uncommon, and a few more extreme cases have been found (Hansen et al. 2005a). Variations in efficacy are primarily a result of the differences in the physical locations (latitude or altitude) of the forcings, which affects the degree to which the forcings can bring climate feedbacks into play, as discussed below.

*Climate sensitivity and climate feedbacks.* Global climate sensitivity is usually defined as the global temperature change that occurs at 'equilibrium', i.e., after the climate system has had a long time to adjust, in response to a specified forcing. The specified forcing is commonly taken to be doubled $CO_2$, thus a forcing of about 4 $W/m^2$.

Climate sensitivity can be evaluated either theoretically, with the help of climate models, or empirically, from the Earth's climate history. In either case, it must be recognized that the climate sensitivity so inferred depends upon what climate variables are fixed as opposed to being allowed to change in response to the climate forcing.

The now famous 1979 National Academy of Sciences study of climate sensitivity (Charney 1979) focused on a case in which atmospheric water vapor, clouds and sea ice are allowed to vary with the climate, but other factors such as ice sheets and the global distribution of vegetation are kept fixed as unchanging boundary conditions. Also long-lived greenhouse gases (GHGs), such as carbon dioxide ($CO_2$), methane ($CH_4$) and nitrous oxide ($N_2O$) are taken as specified boundary conditions or forcings.

In reality all of these boundary conditions can change in response to climate change, becoming either positive climate feedbacks (amplifying the climate change) or negative feedbacks (diminishing the climate change). The choice of feedbacks that were allowed to operate in the Charney (1979) study (water vapor, clouds, sea ice) was in part based on realization that these variables change rapidly, i.e., they are 'fast feedbacks'. Thus if one is interested in climate change on the time scale of decades or longer, these feedbacks must be allowed to operate. Ice sheets and forest cover, on the other hand, might be considered 'slow feedbacks', not expected to change much on decadal time scales. In addition, climate models were not capable of modeling these slower processes at the time of the Charney study.

The Charney (1979) study suggested that equilibrium climate sensitivity was ~3°C (5.4°F) for doubled $CO_2$, with uncertainty at least 50% (1.5°C). Improving climate models continue to yield global climate sensitivity ~3°C for doubled $CO_2$, but uncertainty remains because of the difficulty of accurately simulating clouds.

A more definitive evaluation of climate sensitivity is provided by the Earth's history. With the same choices for the variables specified as forcings, empirical data for climate change over the past 700,000 years yield a climate sensitivity of ¾°C for each $W/m^2$ of forcing, or 3°C for a 4 $W/m^2$ forcing. (see Figure 2 of Reference B). This empirical evaluation of climate sensitivity eliminates concerns about climate models, that they do not realistically simulate cloud processes and they may inadvertently exclude other important processes. The real world climate change contains all processes, including any cloud feedbacks that exist.

*Climate response time.* A practical difficulty with climate change arises from the fact that the climate system does not respond immediately to climate forcings. Figure 1 shows the climate response to a forcing introduced at time t = 0. It requires about 30 years for 50% of the eventual (equilibrium) global warming to be achieved, about 250 years for 75% of the response, and perhaps a millennium for 90% of the surface response.

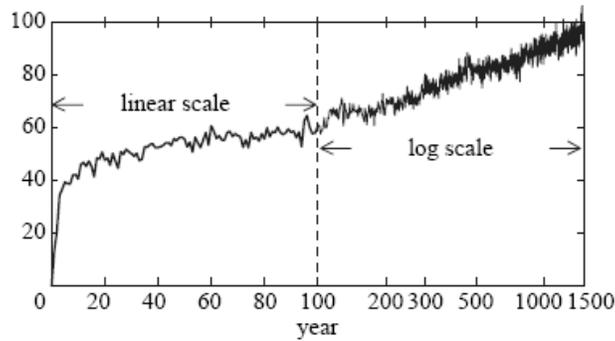

Figure 1. Climate response function (percent of equilibrium response) based on global surface air warming of GISS modelE coupled to Russell ocean model (Reference B).

The exact shape of this response function depends upon the rate of mixing in the ocean, thus upon the realism of the ocean model that is used for its calculation. The response time also depends upon climate sensitivity, the response being slower for higher sensitivity. The reason for this slower response is that climate feedbacks come into play in response to climate change, not in response to the forcing per se, and thus with stronger feedbacks and higher climate sensitivity the response time is longer, indeed, it varies with the square of climate sensitivity (Hansen et al. 1985). The curve in Figure 1 was calculated for sensitivity 3°C for doubled $CO_2$.

This long response time means that even when GHGs stop increasing, there will be additional warming "in the pipeline". Thus we have not yet felt the full climate impact of the gases that have already been added to the atmosphere. This lag effect makes mitigation strategies more arduous.

*Slow climate feedbacks.* The 'Charney', or fast feedback, climate sensitivity is intended to be relevant to decadal time scales. But it is becoming clear that other feedbacks, omitted because they are 'slow' and difficult to deal with, may also be important.

One 'slow' feedback is the poleward movement of forests with global warming. If evergreen forests replace tundra and scrubland vegetation, it makes the surface much darker. Trees are 'designed' to capture photosynthetic radiation efficiently, and thus they can provide a strong positive climate feedback. Forest cover is a powerful positive feedback at Northern Hemisphere high latitudes, and significant changes are already beginning (Zhou et al. 2001; Piao et al. 2006). Although this positive feedback may be partially balanced globally by higher subtropical surface albedo due to increasing desertification, the positive feedback dominates in the regions of possible sea ice and ice sheet tipping points.

Another 'slow' feedback is associated with ice sheets. An ice sheet does not need to disappear for significant feedback to occur: just the change of ice surface albedo (reflectivity) that occurs with increased melt area and melt season duration contributes a large local climate feedback. This feedback occurs in a region where warming is especially important, because of the effect of warming on ice sheet disintegration and sea level rise. Increased areas of surface melt, and lengthening melt season, are observed on both Greenland (Steffen et al. 2004; Fettweis et al. 2007; Tedesco 2007) and West Antarctica (Nghiem et al. 2007).

Still another 'slow' feedback is the effect of warming on emissions of long-lived GHGs from the land or ocean. Melting of tundra in North America and Eurasia is observed to be causing increased ebullition of methane from methane hydrates (Archer 2007; Zimov et al. 2006). In addition, the ability of the ocean to absorb human-made $CO_2$ decreases as the

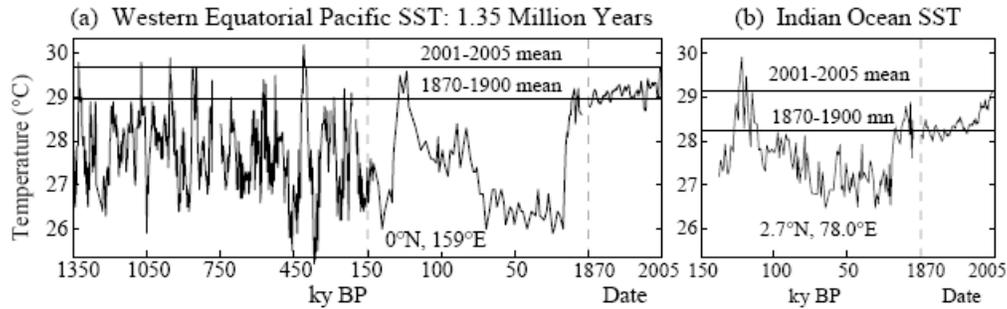

Figure 2. Western Equatorial Pacific (Medina-Elizade and Lea 2005) and Indian Ocean (Saraswat et al. 2005).

emissions increase [Archer, 2005], and there is a possibility that the terrestrial biosphere could even become a source of $CO_2$ [Cox et al. 2000; Jones et al. 2006]

It is apparent that at least some of these 'slow' feedbacks, which are primary causes of the very high climate sensitivity on paleoclimate time scales, as discussed below, are beginning to operate already in response to the strong global warming trend of the past three decades.

B. Earth's History

Civilization developed during the present interglacial period, the Holocene, a period of relatively stable climate, now almost 12,000 years in duration. In this period the Earth has been warm enough to prevent formation of ice sheets in North America or Eurasia, but cool enough to keep stable ice sheets on Greenland and Antarctica. Sea level rose by more than 100 meters between the peak of the last ice age, 20,000 years ago and the Holocene. After sea level finally stabilized, about 7,000 years ago, the first urban centers developed at many points around the globe, perhaps because of the increase in coastal margin productivity that occurred with sea level stabilization and thus the increased availability of high quality food necessary for urban development (Day et al. 2007).

How much warmer does the Earth need to be to destabilize ice sheets and initiate eventual sea level rise of several meters or more? Figures 2 and 3 provide useful indications. With the warming of the past 30 years, key tropical regions are now within 1°C or less of the warmest interglacial periods of the past million years (Figure 2). In the previous interglacial period (about 130,000 years ago), when global mean temperature was not more than about 1°C warmer than today, sea level is estimated to have been $4 \pm 2$ m higher than today (Rostami et al. 2000; Muhs et al. 2002).

It is important to note that the large global climate changes illustrated in Figure 2 are fully accounted for by two mechanisms: changes in the surface albedo of the planet (due to ice sheet area, vegetation distribution, and exposure of continental shelves) and changes in the amount of long-lived greenhouse gases ($CO_2$, $CH_4$, $N_2O$) in the atmosphere. Both the albedo and GHG changes occurred as feedbacks on these long time scales, the principal instigator of the climate changes being changes of the Earth's orbital elements (the tilt of the Earth's spin axis to the orbital plane, the eccentricity of the orbit, and the season of Earth's closest approach to the sun) due to gravitational pull of Jupiter, Saturn and Venus on Earth.

As feedbacks, the albedo and GHG changes tended to lag the climate change by several hundred years. It is probably not coincidental that this lag time is comparable to the ~500 year time scale for ocean turnover. It is important to realize that the response times for the 'slow' feedbacks (greenhouse gas changes, vegetation changes, ice albedo and ice area changes) are much faster than the time scale of the orbital forcing changes.

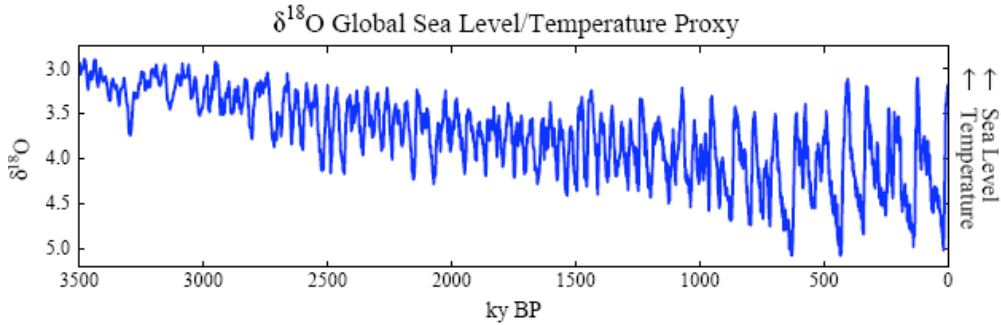

Figure 3. Proxy record of Plio-Pleistocene temperature and ice volume based on oxygen isotope preserved in shells of benthic (deep ocean dwelling) foraminifera (Lisieki and Raymo 2005).

The principal orbital forcing is change of the tilt of the Earth's spin axis, which varies from about 22½° to 24½° at a frequency of about 41,000 years (41 kyr). When the tilt is large it exposes both poles (at 6 month intervals) to increased summer insolation that tends to melt ice sheets, while small tilt allows polar ice sheets to grow. This is the most important orbital forcing, because it has the same sign in the two hemispheres. And this forcing is always present, independent of the eccentricity of the Earth's orbit.

The eccentricity (non-circularity) of the Earth's orbit varies irregularly from about zero (circular orbit) to about 0.06. The time scale of eccentricity changes, as the Earth is tugged by several planets, is not as regular as for tilt changes, but the largest changes are on ~100 kyr time scales. When the eccentricity is significantly different than zero the third orbital parameter comes into play: the season when the Earth is closest to the sun, or, stated differently, the precession of the equinoxes. This precession is the most rapid of the orbital forcings, going through a complete cycle in about 23 kyr.

Eccentricity and precession, working together, cause climate change on ~23 kyr and ~100 kyr periodicities, but the climate forcing is of opposite sign in the two hemispheres, so the net global effect tends to be small, except in special cases as noted below. The eccentricity/ precession forcing functions via its effect on seasonal insolation. Today, for example, the Earth is closest to the sun in January and furthest away in July. This situation favors growth of ice sheets at high latitudes in the Northern Hemisphere, as the relatively warm winters increase atmospheric moisture and snowfall, while the cool summers helps a budding ice sheet to survive.

Thus the natural tendency today, absent humans, would be toward the next ice age, albeit the tendency would not be very strong because the eccentricity of the Earth's orbit is rather small (~0.017). However, another ice age will never occur, unless humans go extinct. Although orbital changes are the 'pacemaker' of the ice ages, the two mechanisms by which the Earth becomes colder in an ice age are reduction of the long-lived GHGs and increase of ice sheet area. But these natural mechanisms are now overwhelmed by human-made emissions, so GHGs are skyrocketing and ice is melting all over the planet. Humans are now in control of global climate, for better or worse. An ice age will never be allowed to occur if humans exist, because it can be prevented by even a 'thimbleful' of CFCs (chlorofluorocarbons), which are easily produced.

But back to the natural world: why did the principal periodicity of ice ages change about one million years ago from 41 kyr to 100 kyr? Figure 3 illustrates this change. $H_2O$ molecules that contain the oxygen isotope $^{18}O$ are heavier and thus move more slowly than $H_2O$ molecules containing the more abundant $^{16}O$. Therefore $H_2O$ molecules with $^{18}O$ evaporate from the ocean less readily. As a result, ice sheets are depleted in $^{18}O$, and as ice sheets grow the proportion of $^{18}O$ in ocean water increases. These changes are recorded in the $^{18}O$ of shells of microscopic marine animals preserved now in oceans sediments.

Figure 3 shows a record of $^{18}O$ in ocean sediments around the world (Lisieki and Raymo 2005). $\delta^{18}O$ ($\delta$ means 'the change of' relative to a standard case) in Figure 3 shows the 41 kyr frequency of global temperature that existed up until about one million years ago when it changed to a frequency of about 100 kyr. Before noting the explanation for this transition, which is quite simple, I need to first note that there are two factors that influence $\delta^{18}O$ significantly: one, already mentioned, is the amount of ice locked in ice sheets (i.e., sea level), and the other is the ocean temperature at the location where the microscopic creatures (benthic, i.e., deep ocean dwelling, foraminifera, whose shells carry the $\delta^{18}O$ record) lived.

The long-term trend in Figure 3 is a consequence of both the ocean becoming colder over that period and more (isotopically light) water being locked in ice sheets on the continents. At the beginning of this period (3.5 million years ago, the middle of the Pliocene epoch) the world was 2-3°C warmer than today and sea level was 25 ± 10 m (80 ± 30 feet) higher (Barrett et al. 1992; Dowsett et al. 1994).

Figure 3 also shows that the amplitude of the glacial-interglacial climate fluctuations increased as the world became colder. This is because the ice/snow albedo feedback becomes larger as the planet becomes colder and has larger area of ice and snow.

The period of the glacial-interglacial swings was ~41 kyr up until one million years ago, because the areas of ice and snow in the two hemispheres were comparable, and thus the effects of eccentricity and precession, opposite in the two hemispheres, tended to largely offset each other in global effect. However, by one million years ago the Earth had become cold enough for a huge ice sheet (called the Laurentide ice sheet) to cover most of Canada, reaching into parts of the United States. A comparable area of ice/snow could not form in the Southern Hemisphere, because at those latitudes (~60°) there is no land in the Southern Hemisphere, but rather roaring east-west ocean currents. This huge asymmetry between the hemispheres allowed the eccentricity/precession effects to become important, so thereafter the global temperature contains signature of all of the ~23, ~41 and ~100 kyr periodicities (Reference B).

The astute reader is probably asking: why was the Earth gradually getting colder, ice area growing, and sea level falling, overall, during the past several million years. The reason, almost surely, was the strong orogeny (mountain building) during the past 10-20 million years. The South American continent has been hitting a rough spot, pushing up the Andes rapidly. It is hard to determine the exact rate, but available evidence indicates, for example, that between 11 and 7 kyr BP (before present) the Andes were rising at a rate of about 1 mm per year, i.e., 1 km per million years (Ghosh et al.). The Himalayas have also been rising rapidly during the past 40 million years (Raymo and Ruddiman 1992), as the Indian plate is crashing into Asia.

Rising mountains increase the rate of weathering of the rocks, and thus the deposition of carbonates on the ocean floor, thus drawing down atmospheric $CO_2$ amount. The precise ice core records of atmospheric $CO_2$ amount go back only about 700,000 years, so we must use much more crude estimates of the atmospheric $CO_2$ content, for example, the stomata of leaves change as atmospheric $CO_2$ changes. From such evidence, it is estimated that the $CO_2$ amount 3½ million years ago was probably in the range 350-450 ppm.

It is apparent that the Earth's history has much to tell us about the degree of atmospheric change that will constitute "danger". I have described some of the empirical information about climate sensitivity and climate feedbacks. There is another vital piece of information in the paleoclimate data that warrants special attention, because it is relevant to what may be the greatest danger that humanity faces with climate change: sea level rise.

One thing that the paleoclimate record shows us is that ice sheet disintegration and sea level rise are usually much more rapid than the opposite process of ice sheet growth and sea level fall. This is reasonable because ice sheet disintegration is a wet process with many positive feedbacks, so it can proceed more rapidly than ice sheet growth, which is limited by the snowfall rate in cold, usually dry, places. At the end of the last ice age sea level rose more than 100 m in less than 10,000 years, thus more than 1 m per century on average. At times during this deglaciation, sea level rose as fast as 4-5 m per century.

If we follow "business-as-usual" GHG emissions, yielding global warming this century of a few degrees Celsius, how long will it take for West Antarctica and Greenland to begin to disintegrate? In the past, an answer to this question has been given based on ice sheet models that were built to try to match paleoclimate records of sea level change. These models tend to require millennia for ice sheets to change by large amounts. It is now reasonably clear that those models were based on a false premise and incomplete physics.

Large sea level changes between glacial and interglacial times typically require several thousand years. However, this is the time scale of the changing forcing, not an inherent response time of the ice sheets. On the contrary, there is no evidence of substantial lag between forcing and ice sheet response. The most rapidly changing paleoclimate forcing has a time scale of 11-12 ky from minimum forcing to maximum forcing, and the changes of sea level are practically coincident with the changes of forcing, suggesting that considerable ice sheet response to forcings can occur within centuries (references B and D above).

C. Current Situation

People are just beginning to notice that climate is changing. Global warming, 1°F in the past 30 years, is much smaller than day to day weather fluctuations or even monthly mean local temperature anomalies. However, the warming is larger over land than over ocean, and the astute observer can note changes that have occurred over the past several decades. Isotherms (lines of a given average temperature) are now moving poleward, in typical land areas, by about 50 km per decade. As this warming continues, or accelerates with "business-as-usual" GHG emissions, it will begin to have dramatic effects, as discussed in the next section.

To understand the urgency of addressing the global warming problem, it is necessary to recognize a critical distinction that exists among pollution problems arising in the fossil-fuel-driven industrial revolution. When the industrial revolution began in Britain it was powered first by coal, the most abundant of the fossil fuels. Later discoveries of oil and gas, which are more mobile and convenient fossil fuels, provided energy sources that helped power the developed world to ever greater productivity and living standards.

We did not face up to the dark side of the industrial revolution until it was thrust in our face. London choked on smog. A river in the United States burned. Forests were damaged by acid rain. Fish died in many lakes. These problems were traced to pollutants from fossil fuels.

We have solved or are solving those pollution problems, at least in developed countries. But we did not address them until they hit us with full force. That approach, to wait and see and fix the problems post facto, unfortunately, will not work in the case of global climate change. On the contrary, the inertia of the climate system, the fact that much of the climate change due to gases already in the air is still 'in the pipeline', and the time required for economically-sensible phase-out of existing technologies together have a profound implication. They imply that ignoring the climate problem at this time, for even another decade, would serve to lock in future catastrophic climatic change and impacts that will unfold during the remainder of this century and beyond (references A and B).

Yet this is not a reason for gloom and doom. On the contrary, there are many bright sides to the conclusion that the 'dangerous' level of $CO_2$ is no more than 450 ppm, and likely much less than that. It means that we, humanity, are forced to find a way to limit atmospheric $CO_2$ more stringently than has generally been assumed. In so doing, many consequences of high $CO_2$ that were considered inevitable can be avoided. We will be able to avoid acidification of the ocean with its destruction of coral reefs and other ocean life, retain Arctic ice, limit species extinctions, prevent the U.S. West from become intolerably hot, and avoid other undesirable consequences of large global warming.

It is becoming clear that we must make a choice. We can resolve to move rapidly to the next phase of the industrial revolution, and in so doing help restore wonders of the natural world, of creation, while maintaining and expanding benefits of advanced technology. Or we can continue to ignore the problem, creating a different planet, with eventual chaos for much of humanity as well as the other creatures on the planet.

## 4. Metrics for Dangerous Climate Change

I have argued elsewhere (Hansen 2006) that ice sheet disintegration and extermination of species deserve high priority as metrics for dangerous climate change, because, for all practical purposes, these consequences are irreversible. Regional climate change also has great impacts on humanity.

A. Sea Level Rise

The sharpest criterion for defining dangerous climate change is probably maintenance of long-term sea level close to the present level (reference A), as about one billion people live within 25 m elevation of today's sea level. These areas (Figure 4) include many East Coast U.S. cities, almost all of Bangladesh, and areas occupied by more than 250 million people in China.

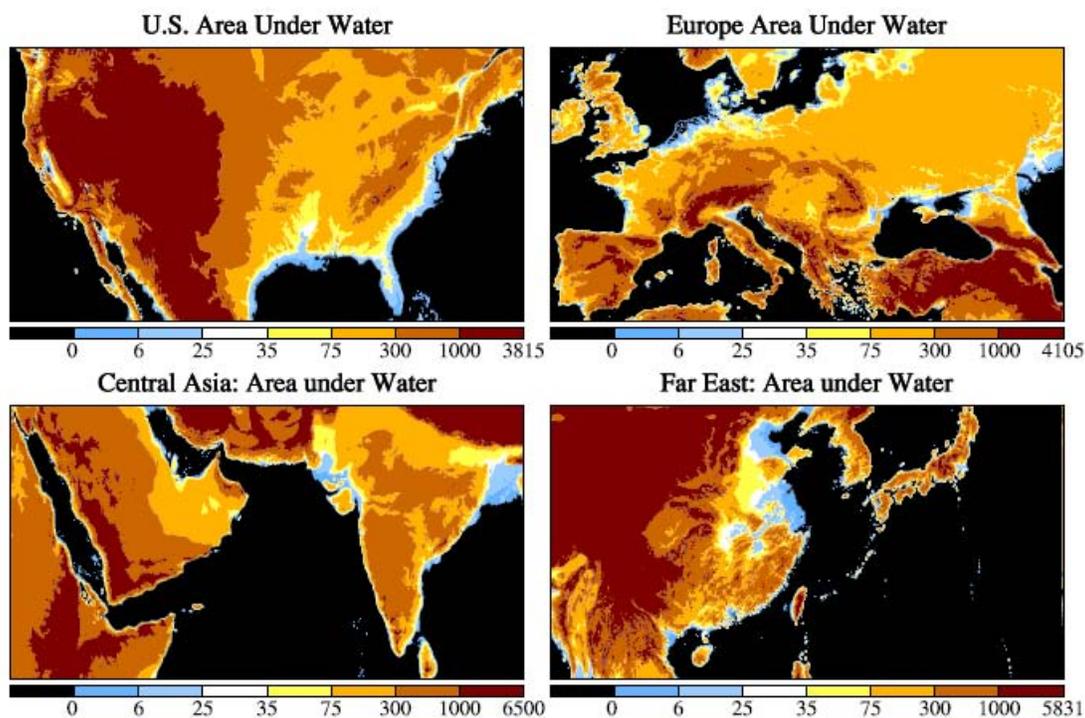

Figure 4. Areas under water for specified sea level increases. Blue regions would be expected to eventually be under water with global warming as great as that of the middle Pliocene (2-3°C).

Sea level change has received less attention in the past than it warrants, because of a common presumption that ice sheet inertia prevents substantial changes on time scales shorter than millennia. Closer inspection of paleoclimate data calls that assumption into question, and increasingly rapid changes on West Antarctica and Greenland, observed by satellite and in the field during the past few years, are alarming. West Antarctica is of particular concern, because, as a marine based ice sheet, global warming attacks it from both below and above.

Sea level is already rising at a rate of 3.5 cm per decade (more than one foot per century) and the rate is accelerating. It is impossible to say at exactly what level of global warming the rate of sea level rise could accelerate to meters per century, because ice sheet disintegration is a very non-linear process in which changes can occur suddenly. But paleoclimate data suggests that we are not far from such a level of global warming.

B. Extermination of Species

Climate change is emerging while the state of the wild is stressed by other forces. Pressures include destruction of habitat, hunting and resource use, pollution, and introduction of exotic competing species. Climate effects are magnified by these other stresses, including human-caused fragmentation of ecosystems. As a result, continued business-as-usual greenhouse gas emissions threaten many ecosystems and their species, which together form the fabric of life on Earth and provide a wide range of services to humanity.

Animals and plants migrate as climate changes, but their potential escape routes may be limited by geography or human-made obstacles. Polar species can be pushed off the planet, as they have no place else to go. In Antarctica, Adelie and emperor penguins are in decline, as shrinking sea ice has reduced the abundance of krill, the penguins shrimp-like food source (Gross 2005).

Arctic polar bears are also feeling the pressure of melting sea ice. Polar bears hunt seals on the sea ice and fast in the summer, when the ice retreats from shore. As ice is receding earlier, some populations of bears in Canada have declined about 20%, with the weight of females and the number of surviving cubs decreasing a similar amount.

The apparent good news is that the U.S. Fish and Wildlife Service is considering whether it will protect polar bears under the Endangered Species Act (Pennisi 2007). I say apparent, because the announcement was made only after the Fish and Wildlife Service was taken to court for failure to act. And connection of polar bear plight to greenhouse gas emissions has been drawn only by those bringing suit, not by the government.

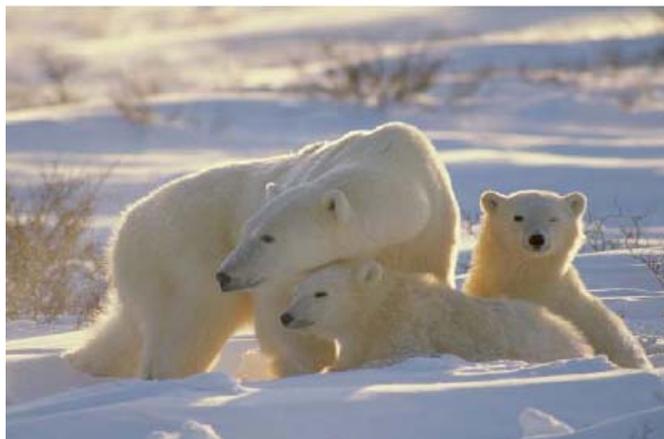

Figure 5. Polar bear are threatened by global warming. The weight of females and number of cubs in some populations have decreased about 20 percent. (Image Credit: Paul Burke, First People)

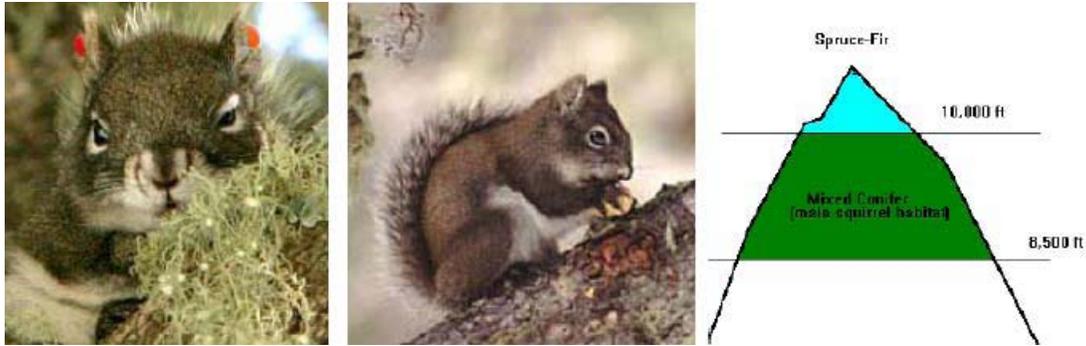

Figure 6. Mount Graham Red Squirrel survives on a green 'island in the sky', surrounded by desert. (Credits: PHOTOSMITH, 2004, Claire Zugmeyer and Bruce Walsh, University of Arizona.)

     The apparent good news is that the U.S. Fish and Wildlife Service is considering whether it will protect polar bears under the Endangered Species Act (Pennisi 2007). I say apparent, because the announcement was made only after the Fish and Wildlife Service was taken to court for failure to act. And connection of polar bear plight to greenhouse gas emissions has been drawn only by those bringing suit, not by the government.

     Life in alpine regions, including the biologically diverse slopes leading to the mountains, is similarly in danger of being pushed off the planet. As a given temperature range moves up the mountain the area with those climatic conditions becomes smaller and rockier, and the air thinner. The resulting struggle for life is already becoming apparent in the southwest United States, where the effects are hastened by intensifying drought and fire.

     The Mount Graham red squirrel survives now on a single Arizona mountain, one of the 'islands in the sky' in the American Southwest. These 'islands' are green regions scattered on mountains in the desert. Stresses on this species include introduction of a grey squirrel that raids the food middens built by the red squirrel. Classified as endangered, the Graham red squirrel population rebounded to over 500 by 1999 (Jordan 2006), but has since declined to between 100 and 200 (Egan 2007). Loss of the red squirrel will alter the forest, as its middens are a source of food and habitat for chipmunks, voles and mice.

     The new stress driving down Graham red squirrel numbers, perhaps toward extinction, is climatic: increased heat, drought and fires. Heat-stressed forests are vulnerable to prolonged beetle infestation and catastrophic fires. Rainfall still occurs, and when it does it can be substantial because warmer air holds more water. But dry periods are more intense and resulting forest fires burn hotter, thus leaving an almost-lifeless 'scorched earth' so devastated that lower reaches of the forest cannot recover, becoming part of the desert below.

     Might the Graham red squirrel be 'saved' by transplantation to a higher mountain, where it could compete for a niche? One difficulty would be the 'tangled bank' of interactions that has evolved among species (Montoya et al. 2006). What is the prospect that humans can understand, let alone reproduce, all the complex interactions that create ecological stability? 'Assisted migration' thus poses threats to other species (Zimmer 2007), as well as uncertain prospects for those that are transported.

     The underlying cause of the climatic threat to the Graham Red Squirrel, and millions of other species, is continued 'business-as-usual' increase of fossil fuel use. The best chance for all species, including humans, is a conscious choice by the latter species to pursue an alternative energy scenario, one leading to stabilization of climate.

C. Regional Climate Change

Regional climate changes due to global warming may have the greatest impact on humans in the near-term. Changes of the hydrologic cycle are of special concern. An expansion and intensification of subtropical dry conditions occurs consistently in climate model simulations of global warming. Practical impacts include increased drought and forest fires in regions such as the Western United States, Mediterranean, Australia, and parts of Africa. Paleoclimate data provide further evidence of increased drought in the Western United States accompanying warmer climate.

It is difficult to specify a threshold for 'dangerous' regional effects. Business-as-usual greenhouse gas scenarios yield regional temperature changes 5-10 times the standard deviation of regional variations in the past century, which is argued (reference A) to be prima facie evidence of dangerous change. Regional climate change, as well as sea level and species, would be protected by stabilizing global warming near its current level.

## 5. Four-Point Strategy to Stabilize Climate

The evidence that the Earth's atmosphere is already close to having a dangerous level of greenhouse gases is no cause for despair. It implies only that we must resolve to move promptly to the next phase of the industrial revolution. In doing so, we can help restore wonders of the natural world, of creation, while maintaining and expanding benefits of advanced technology.

A framework for actions that are needed becomes apparent upon review of basic fossil fuel facts. Figure 7a shows estimated amounts of $CO_2$ in each fossil fuel reservoir: oil, gas, coal and unconventional fossil fuels (tar sands, tar shale, heavy oil, methane hydrates). A significant fraction of oil and gas has already been used (dark portion of bar graph). Proven and anticipated reserves are based on Energy Information Administration estimates. Other experts estimate higher or lower reserves, but the uncertainties do not alter our conclusions.

Data on fossil fuel reservoirs must be combined with knowledge about the 'carbon cycle'. The ocean quickly takes up a fraction of fossil fuel $CO_2$ emissions, but uptake slows as $CO_2$ added to the ocean exerts a 'back pressure' on the atmosphere. Further uptake then depends upon mixing of $CO_2$ into the deep ocean and ultimately upon removal of $CO_2$ from the ocean via formation of carbonate sediments. As a result, one-third of fossil fuel $CO_2$ emission remains in the air after 100 years and one-quarter still remains after 500 years (Fig. 9 of reference A).

One conclusion from these fossil fuel facts is that readily available oil and gas resources alone will take atmospheric $CO_2$ to the neighborhood of 450 ppm. Coal and unconventional fossil fuels could take atmospheric $CO_2$ to far greater levels. These carbon reservoirs are an important boundary condition in framing solutions to the climate crisis.

A second boundary condition is the Earth's energy imbalance, which defines the 'momentum' of the climate system. Creation of 'a different planet', with an ice-free Arctic and eventual disintegration of ice sheets, can be averted only if planetary energy balance is restored at an acceptable global temperature, i.e., one that avoids these catastrophic changes. Estimates of permissible additional warming must be refined as knowledge advances and technology improves, but the upshot of crystallizing science is that the 'safe' global temperature level is, at most, about 1°C greater than year 2000 temperature. It may be less, indeed, I suspect that it is less, but that does not alter the nature of the actions that are needed.

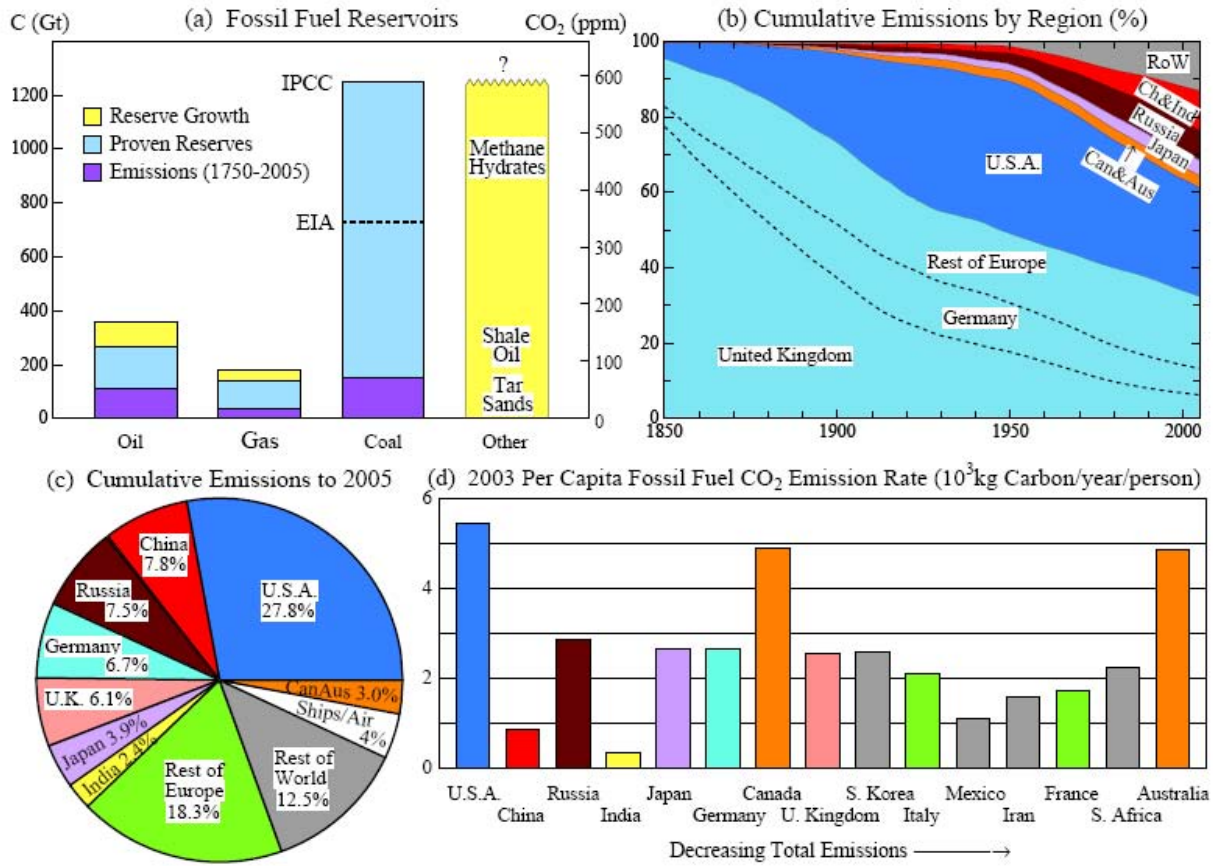

Figure 7. (a) $CO_2$ contained in fossil fuel reservoirs, the dark area being the portion already used, (b, c) cumulative fossil fuel $CO_2$ emissions by different countries as a percent of global total, (d) per capita emissions for the ten largest emitters of fossil fuel $CO_2$ (data from Marland 2006).

A 1°C limit on added global warming implies a $CO_2$ ceiling of about 450 ppm (reference A). There is some 'play' in the $CO_2$ ceiling due to other human-made climate forcings that cause warming, especially $CH_4$, $N_2O$, and 'black soot'. The 'alternative scenario' (Hansen et al. 2000), designed to keep additional warming under 1°C, has $CO_2$ peaking at 475 ppm via an assumed large reduction of $CH_4$. However, human-made sulfate aerosols, which have a cooling effect, are likely to decrease and tend to offset reductions of positive non-$CO_2$ forcings. Therefore 450 ppm is a good first estimate of the maximum allowable $CO_2$. But if recent mass loss in Antarctica is the beginning of a growing trend, it is likely that even 450 ppm is excessive and dangerous.

The low limit on allowable $CO_2$ has a bright side. Such a limit requires changes to our energy systems that would do more than solve the sea level problem. They would leave ice in the Arctic and avoid dramatic regional climate changes. Air pollutants produced by fossil fuels, especially soot and low level ozone, also would be reduced, thus restoring a more pristine, healthy planet. With climate little different than at present, most species could survive.

An outline of the strategy that humanity must follow to avoid dangerous climate change emerges from the above boundary conditions. It is a four-point strategy (following table).

**Methods to Reduce CO$_2$ Emissions**
1. Energy Efficiency & Conservation
   More Efficient Technology
   Life Style Changes
2. Renewable & CO$_2$-Free Energy
   Hydro
   Solar, Wind, Geothermal
   Nuclear
3. CO$_2$ Capture & Sequestration
   → No Silver Bullet
   → All Three are Essential

**Outline of Solution**
1. Coal only in Powerplants w Sequestration
   Phase-out old technology. Timetable TBD
2. Stretch Conventional Oil & Gas
   Via Incentives (Carbon tax) & Standards
   No Unconventional F.F. (Tar Shale, etc.)
3. Reduce non-CO$_2$ Climate Forcings
   Methane, Black Soot, Nitrous Oxide
4. Draw Down Atmospheric CO$_2$
   Agricultural & Forestry Practices
   Biofuel-Powered Power-Plants

A. Coal and Unconventional Fossil Fuels

First, coal and unconventional fossil fuels must be used only with carbon capture and sequestration. Existing coal-fired power plants must be phased out over the next few decades. This is the primary requirement for avoiding 'a different planet'.

It is impractical to prevent use of most of the easily extractable oil and resulting CO$_2$ emission from small mobile sources. This makes it imperative to use the huge coal resource in a way such that the CO$_2$ can be captured, and, indeed, the logical use of coal is in power plants. This imperative is a consequence of the low limit on allowable atmospheric CO$_2$, the size of the fossil fuel reservoirs, and the fact that a substantial fraction of emitted CO$_2$, if it is not captured, will remain in the air for 'an eternity', more than 500 years.

Thus the most effective action at this time, I believe, is to avoid construction of additional coal-fired power plants without CO$_2$ capture capability. Given that many governments around the world, not only in the United States, China and India, fail to appreciate this situation, it is important that citizens draw attention to the issue. The most effective action that people can take to stem global warming, I suspect, is to help prevent construction of coal-fired power plants until sequestration technology is ready. In the interim, there is sufficient potential in energy efficiency and renewable energies to handle increased energy needs.

B. Stretching Oil and Gas with a Carbon Tax

Oil and gas must be 'stretched' so as to cover needs for mobile fuels during the transition period to the next phase of the industrial era 'beyond petroleum'. This 'stretching', almost surely, requires a continually rising price on carbon emissions. Innovations in energy efficiency and renewable energy will be unleashed if industry realizes that this rising price is certain. Efficiency standards, for vehicles, buildings, appliances, and lighting are needed, as well as a carbon price. The rising carbon tax will also help avert the threat of emissions from unconventional fossil fuels, such as tar shale.

There needs to be recognition that vehicle fuels must transition to a future with no substantial carbon emissions. Limited land availability makes biofuels an unlikely long-term solution for vehicle propulsion, at least as the principal fuel. Electric propulsion allows increased options for primary energy and has other desirable attributes, including the potential to minimize air pollution. As battery technology and vehicle efficiencies improve, electric and hybrid-electric propulsion appear to have great potential. However, it is unhelpful for governments to specify technologies; it is better for 'winners' to be chosen via the market place and a rising tax on carbon emissions. Tax rates should be set so as to account for estimated full life cycle emissions of the fuel source, which would effect, e.g., the nature and viability of any potential ethanol contributions to energy needs.

C. Non-$CO_2$ Climate Forcings

Reduction of non-$CO_2$ forcings can be a big help in achieving the climate forcings needed to keep climate change within given bounds (Hansen et al. 2000). The sum of the climate forcings due to human-made $CH_4$, $N_2O$, $O_3$, and CFCs is comparable to the forcing by $CO_2$. The non-$CO_2$ forcings have received little attention, perhaps in part because their growth rates have slowed. However, this does not reduce the value of further reductions of their emissions. It may even be possible to achieve absolute reductions in the atmospheric amounts of these gases, thus balancing part of the greenhouse effect of increasing $CO_2$. Indeed, we have argued (reference A) that avoidance of dangerous climate change probably requires a 'full court press' to both slow emissions of $CO_2$ and obtain an absolute reduction of non-$CO_2$ forcings.

Black carbon (BC) aerosols, or 'black soot', also contributes to global warming by warming the lower atmosphere and increasing the melting of snow and ice (Hansen 2000; Hansen et al. 2005; Flanner et al. 2007). In part because of BC, the non-$CO_2$ forcings are especially effective in the Arctic (reference A). Thus we suggest (reference A) that it is still possible to retain Arctic ice via an absolute reduction of non-$CO_2$ forcings together with minimization of further $CO_2$ increase.

Reduction of non-$CO_2$ forcings has benefits for human health and agriculture (Hansen 2002; West et al. 2005), as well as for climate. Because of the magnitude of air pollution problems in the developing world, a combined emphasis on reducing $CO_2$ and non-$CO_2$ forcings, with technological assistance to developing countries, has the potential to aid attainment of global cooperation in limiting climate change.

D. Drawing Down Atmospheric $CO_2$

Because $CO_2$ is already near the dangerous level, actions will be needed to 'draw down' atmospheric $CO_2$. Farming and forestry practices that enhance carbon retention and storage in the soil and biosphere should be supported (McCarl et al. 2007).

In addition, burning biofuels in power plants with carbon capture and sequestration could draw down atmospheric $CO_2$ (Hansen 2007), in effect putting anthropogenic $CO_2$ back underground where it came from. $CO_2$ sequestered beneath ocean sediments is inherently stable (House et al. 2006). Other safe geologic sites may also be available.

This use of biofuels in a power plant, which would draw down atmospheric $CO_2$, should be contrasted with use of corn-based ethanol to power vehicles. The latter process still results in large increases of atmospheric $CO_2$, increases food prices worldwide, and results in deforestation and poor agricultural practices as greater land area is pressed into service. In contrast, a sustainable practice would be to use native grasses harvested with non-till practices and other cellulosic fibers as principal feedstocks in producing biofuels for power plants.

Limited land availability makes biofuels an unlikely long-term solution for vehicle propulsion, at least as the principal fuel. As noted above, electric propulsion allows increased options for primary energy and has other desirable attributes, including the potential to minimize air pollution. As battery technology and vehicle efficiencies improve, electric and hybrid-electric propulsion appear to have great potential, but there is unhelpful for governments to attempt to define specific 'winning' technologies.